\documentclass[sigconf,nonacm]{acmart}
\usepackage{graphicx}
\usepackage{subfig}
\usepackage{xparse}
\usepackage{enumitem}
\usepackage{multirow}
\usepackage{tcolorbox}
\usepackage{xspace}
\usepackage[T1]{fontenc}
\usepackage[utf8]{inputenc}

\tcbuselibrary{listings}
\newtcolorbox[auto counter, number within=section]{cuteconversationblue}[2][]{colback=blue!5!white, colframe=blue!75!white, colbacktitle=blue!10!white, coltitle=black, title=#2,#1}
\newtcolorbox[auto counter, number within=section]{cuteconversationred}[2][]{colback=red!5!white, colframe=red!75!white, colbacktitle=red!10!white, coltitle=black, title=#2,#1}
\newtcolorbox[auto counter, number within=section]{cuteconversationcyan}[2][]{colback=cyan!5!white, colframe=cyan!75!white, colbacktitle=cyan!10!white, coltitle=black, title=#2,#1}

\setlist[description]{
  font={\sffamily\bfseries},
  labelsep=0pt,
  labelwidth=\transcriptlen,
  leftmargin=\transcriptlen,
}
\newlength{\transcriptlen}
\NewDocumentCommand {\setspeaker} { mo } {%
  \IfNoValueTF{#2}
  {\expandafter\newcommand\csname#1\endcsname{\item[#1:]}}%
  {\expandafter\newcommand\csname#1\endcsname{\item[#2:]}}%
  \IfNoValueTF{#2}
  {\settowidth{\transcriptlen}{#1}}%
  {\settowidth{\transcriptlen}{#2}}%
}
\setspeaker{Chad}
\setspeaker{Alex}
\setspeaker{ChatGPT}
\newif\ifcomment\commentfalse
\commenttrue

\ifcomment
\newcommand\sadra[1]{\textcolor{magenta}{[For/From Sadra:] \textit{#1}}}
\newcommand\sepand[1]{\textcolor{blue}{[For/From Sepand:] \textit{#1}}}
\else 
\newcommand\sadra[1]{}
\newcommand\sepand[1]{}
\fi

\newcommand{\nafas}{\emph{Nafas}\xspace}
\title[Nafas: Breathing Gymnastics Application]{\nafas: Breathing Gymnastics Application}

\author{Sadra Sabouri}
\affiliation{ 
  \institution{Open Science Laboratory (OpenSciLab)}
  \institution{University of Southern California}
  \country{United States of America}
}
\email{sadra@openscilab.com}
\email{sabourih@usc.edu}

\author{Sepand Haghighi}
\affiliation{ 
  \institution{Open Science Laboratory (OpenSciLab)}
   \country{United States of America}
}
\email{sepand@openscilab.com}

\begin{document}

\begin{abstract}
    \setlength{\leftskip}{5mm} % Adjust left margin
    \setlength{\rightskip}{5mm} % Adjust right margin
    Long sessions of computer use introduce physical and mental health risks, particularly for programmers and intensive computer users. Breathing exercises can improve focus, reduce stress, and overall well-being. However, existing tools for such practices are often app-based, requiring users to leave their workspace. In this technical report, we introduce \nafas, a command-line interface (CLI) application designed specifically for computer users, enabling them to perform breathing exercises directly within the terminal. \nafas offers structured breathing programs with various levels tailored to the needs of busy developers and other intensive computer users.
\end{abstract}

\maketitle

\section*{Disclaimer}
The breathing program techniques in this application are not intended to diagnose, treat, cure, or prevent any medical condition.
These programs are designed as practical exercises developed by general observations and evaluations to promote relaxation, focus, and well-being in healthy individuals.
For curing intents or medical treatments, we recommend referring to established medical literature and consulting healthcare professionals.

\section{Introduction}
Using computers in prolonged sessions become a defining characteristic of modern work and leisure, especially for programmers and other intensive users who spend hours in front of screens daily.
This extended screen time can lead to various health issues~\cite{ellahi2011computer}, including physical strain~\cite{taib2016effect}, mental fatigue~\cite{van2003impact}, and increased stress~\cite{walz2012stress}.
Physical symptoms, such as eye strain~\cite{kaur2022digital} and musculoskeletal discomfort~\cite{blatter2002duration} are common, while mental challenges include heightened anxiety~\cite{kim2016computer}, reduced attention span~\cite{espiritu2016early}, and susceptibility to burnout~\cite{salanova2002self}.
The cumulative effect of these issues is reduced productivity in work and quality of life.

Breathing exercises are increasingly recognized as effective ways to counteract these health challenges, especially for extensive computer users~\cite{tabor2024designing,fernandes2021exploring,araya2021benefits}.
Slow and controlled breathing has been shown to calm the nervous system, enhance heart rate variability, improve respiratory efficiency, and reduce stress and anxiety responses \cite{russo2017physiological, zaccaro2018breath}.
Regular breathing exercises can also enhance focus and improve emotional regulation, contributing to an overall sense of well-being \cite{analayo2019meditation}.
Studies show that breath control can mitigate the impacts of stress and anxiety~\cite{novaes2020effects}, making it particularly beneficial for high-intensity tasks like programming \cite{valenza2014effectiveness, jerath2015self}. 

While various applications support breathing exercises, they are often app-based, requiring users to leave their primary workspace to engage with them. For many users, especially developers accustomed to command-line workflows, this shift can disrupt productivity and deter regular use. Tools that integrate directly into the user’s environment, the Command-Line-Interface (CLI) application, in this case, offer a more seamless experience, allowing users to perform breathing exercises without interrupting their workflow.

In this technical report, we introduce \nafas\footnote{\url{https://github.com/sepandhaghighi/nafas}}, which means ``breath'' in Farsi, a CLI application designed to address these needs by providing in-terminal access to breathing exercises.
Unlike traditional well-being applications, \nafas allows computer users to use structured breathing programs without leaving their working environment.
\nafas promotes mental and physical well-being in a format that complements, rather than disrupts, the user’s workflow.
\nafas has gained significant popularity since its release in August 2020. As of writing this technical report, with eight versions released, it has received 150 stars and has been downloaded 16,000 times.
This technical report is based on Nafas version 0.8.

\section{Related Work}

In this section, we review related works in breathing techniques for health, well-being tools tailored for developers, and existing approaches to integrate mindfulness in the workspace. By examining the limitations of current tools, we identify gaps that \nafas seeks to address as a CLI-based breathing application.

\textbf{Health Concerns and Solutions for Computer Users.}
Extensive computer use, especially among developers, can lead to various health issues like Computer
Vision Syndrome (CVS)~\cite{akinbinu2014impact} and Carpal Tunnel Syndrome~\cite{andersen2003computer}.
To address physical strain, ergonomic tools like adjustable chairs and desks~\cite{chandra2009ergonomics}, along with regular breaks~\cite{henning1997frequent}, are commonly recommended.
For mental well-being, mindfulness techniques such as slow breathing~\cite{lan2021slow} exercises can help reduce stress and improve focus.

\textbf{Slow Breathing and Health Benefits.}
Slow breathing, defined as a rate of 4 to 10 breaths per minute (0.07–0.16 Hz) compared to the typical 10–20 breaths per minute (0.16–0.33 Hz)~\cite{russo2017physiological}, provides numerous physiological and psychological benefits~\cite{campanelli2020pranayamas}.
It improves respiratory efficiency by increasing lung volume, optimizing airflow, and reducing dead space, which enhances oxygen delivery to the bloodstream~\cite{russo2017physiological}.
Cardiovascular benefits include increased heart rate variability (HRV), better synchronization of blood pressure and heartbeat rhythms, improved circulation, and potentially lower blood pressure~\cite{russo2017physiological}. Additionally, slow breathing promotes coordination between the heart and lungs, improving gas exchange while reducing the heart's workload and regulating blood pressure through aligned breathing and heart rhythms~\cite{russo2017physiological}.
From a nervous system perspective, slow breathing strengthens vagal nerve activity, shifting the body to a relaxed parasympathetic state and balancing the stress-relaxation response, enhancing adaptability and stability~\cite{russo2017physiological, zaccaro2018breath}.
Psychologically, slow breathing increases relaxation, comfort, and alertness while reducing negative emotional states such as anxiety~\cite{novaes2020effects}, depression, and anger~\cite{zaccaro2018breath}. 

\textbf{Well-being Tools for Developers.}
Several well-being tools cater specifically to developers, focusing on practices like mindful breaks, physical exercise reminders, and productivity monitoring.
Tools such as Stretchly~\cite{strechly} and Workrave~\cite{workrave} are designed to encourage short, periodic breaks to prevent burnout and physical strain. These applications operate in a graphical environment and often interrupt the user with pop-ups or reminders, which can disrupt the workflow or become obtrusive.

\nafas fills this gap by offering a non-intrusive, CLI-based wellness tool that aligns with developers' working habits. Unlike app-based or web-based wellness tools that require leaving the workspace, \textit{Nafas} allows users to perform breathing exercises directly in the terminal, supporting both physical and mental health in a way that integrates seamlessly with their work routines.

\section{Structure}
% Define column width parameters
\newcommand{\programColWidth}{1.5cm}
\newcommand{\descriptionColWidth}{10cm}
\newcommand{\levelColWidth}{1cm}
\newcommand{\ratiosColWidth}{1.5cm}
\newcommand{\unitColWidth}{1.5cm}
\newcommand{\cyclesColWidth}{1cm}

\begin{table*}[h!]
    \centering
        \caption{Overview of \textit{Nafas} Breathing Programs}
    \resizebox{\textwidth}{!}{%
    \begin{tabular}{|p{\programColWidth}|p{\descriptionColWidth}|p{\levelColWidth}|p{\ratiosColWidth}|p{\unitColWidth}|p{\cyclesColWidth}|}
    \hline
    \textbf{Program} & \textbf{Description} & \textbf{Level} & \textbf{Ratios} & \textbf{Unit} & \textbf{Cycles} \\
     &  &  & \textbf{(I:R:E:S)} & \textbf{(Seconds)} & \\
    \hline
    \multirow{3}{\programColWidth}{Clear Mind~\cite{albul_2014_prana}} & \multirow{3}{\descriptionColWidth}{Short program to reduce mental fog and improve focus. Useful for developers during coding breaks. Beneficial for those needing quick mental refreshment.} & B & 1:0:3:0 & 3 & 35 \\
    & & M & 1:0:4:0 & 3 & 28 \\
    & & A & 1:0:5:0 & 3 & 24 \\
    \hline
    \multirow{3}{\programColWidth}{Relax1~\cite{albul_2014_prana}} & \multirow{3}{\descriptionColWidth}{Relaxation-focused program to calm nerves. Ideal for use after intense coding sessions. Useful for programmers managing stress.} & B & 1:0:2:2 & 3 & 28 \\
    & & M & 1:0:2:3 & 3 & 24 \\
    & & A & 1:0:2:4 & 3 & 22 \\
    \hline
    \multirow{3}{\programColWidth}{Relax2~\cite{weil_2014_dr}} & \multirow{3}{\descriptionColWidth}{Enhanced relaxation program with a focus on breathing depth. Suitable for unwinding after long work sessions.} & B & 4:7:8:0 & 1 & 4 \\
    & & M & 4:7:8:0 & 1 & 8 \\
    & & A & 4:7:8:0 & 1 & 12 \\
    \hline
    \multirow{3}{\programColWidth}{Relax3~\cite{humangivensinstitute_2017_711}} & \multirow{3}{\descriptionColWidth}{Deeper relaxation exercise for stress relief and a more meditative state. Ideal for unwinding after complex projects.} & B & 7:0:11:0 & 1 & 15 \\
    & & M & 7:0:11:0 & 1 & 20 \\
    & & A & 7:0:11:0 & 1 & 24 \\
    \hline
    \multirow{3}{\programColWidth}{Calming1~\cite{albul_2014_prana}} & \multirow{3}{\descriptionColWidth}{Soothing breathing pattern, helpful for reducing anxiety. Beneficial for long working hours and maintaining calm.} & B & 1:2:1:2 & 3 & 24 \\
    & & M & 1:3:1:3 & 3 & 22 \\
    & & A & 1:4:1:4 & 3 & 20 \\
    \hline
    \multirow{3}{\programColWidth}{Calming2~\cite{a2015_using}} & \multirow{3}{\descriptionColWidth}{An extended version of the calming routine, providing a longer period of relaxation and anxiety reduction. Useful for balancing high-stress situations.} & B & 5:0:5:5 & 1 & 4 \\
    & & M & 5:0:5:5 & 1 & 6 \\
    & & A & 5:0:5:5 & 1 & 8 \\
    \hline
    \multirow{3}{\programColWidth}{Power~\cite{albul_2014_prana}} & \multirow{3}{\descriptionColWidth}{Energizing breathing, designed for boosting energy and alertness. Suitable for users feeling fatigue.} & B & 1:2:2:0 & 3 & 28 \\
    & & M & 1:3:2:0 & 3 & 24 \\
    & & A & 1:4:2:0 & 3 & 20 \\
    \hline
    \multirow{3}{\programColWidth}{Harmony~\cite{albul_2014_prana}} & \multirow{3}{\descriptionColWidth}{Balancing exercise to promote relaxation and focus. Ideal for programmers looking to maintain concentration.} & B & 1:3:2:1 & 3 & 20 \\
    & & M & 1:4:2:1 & 3 & 18 \\
    & & A & 1:5:2:1 & 3 & 16 \\
    \hline
    \multirow{3}{\programColWidth}{Anti-Stress~\cite{albul_2014_prana}} & \multirow{3}{\descriptionColWidth}{Specialized for stress relief, particularly useful during high-stress tasks or debugging sessions.} & B & 3:0:0.66:0 & 3 & 20 \\
    & & M & 4:0:0.66:0 & 3 & 17 \\
    & & A & 5:0:0.66:0 & 3 & 14 \\
    \hline
    \multirow{3}{\programColWidth}{Anti-Appetite~\cite{albul_2014_prana}} & \multirow{3}{\descriptionColWidth}{Breathing program to manage cravings and avoid unnecessary snacking during work. Beneficial for users seeking appetite control.} & B & 5:0:5:5 & 1 & 40 \\
    & & M & 6:0:5:5 & 1 & 38 \\
    & & A & 7:0:5:5 & 1 & 36 \\
    \hline
    \multirow{3}{\programColWidth}{Cigarette Replace~\cite{albul_2014_prana}} & \multirow{3}{\descriptionColWidth}{Alternative to smoking breaks, encouraging mindful breathing. Suitable for smokers looking to reduce habit triggers.} & B & 2:1.1:2.2:0.8 & 2 & 23 \\
    & & M & 3:1.1:2.2:0.8 & 2 & 21 \\
    & & A & 4:1.1:2.2:0.8 & 2 & 19 \\
    \hline
    \multirow{3}{\programColWidth}{Decision-Making~\cite{mithustoroni_2019_this}} & \multirow{3}{\descriptionColWidth}{Breathing exercise to improve focus before making critical decisions. Useful for developers or managers.} & B & 5:2:7:0 & 1 & 6 \\
    & & M & 5:2:7:0 & 1 & 10 \\
    & & A & 5:2:7:0 & 1 & 14 \\
    \hline
    \multirow{3}{\programColWidth}{Balancing~\cite{burgin_2020_pranayama}} & \multirow{3}{\descriptionColWidth}{Designed for achieving inner balance. Useful for users needing a brief grounding exercise during hectic schedules.} & B & 6:0:6:0 & 1 & 6 \\
    & & M & 8:1:8:1 & 1 & 8 \\
    & & A & 6:2:6:2 & 1 & 10 \\
    \hline
    \multicolumn{6}{l}{
            I: Inhale, R: Retain, E: Exhale, S: Sustain, B: Beginner, M: Medium, A: Advanced}
    \end{tabular}%
    }
    \label{tab:nafas_programs}
\end{table*}

\begin{figure}
    \centering
    \includegraphics[width=\linewidth]{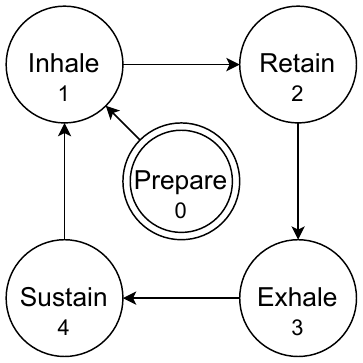}
    \caption{\nafas Programs' Cycle. Each program starts with a preparation phase and goes through several inhales and exhalations with pauses as it retains and sustains.}
    \label{fig:nafas-cycles}
\end{figure}

In this section, we describe parts of \nafas as a CLI-based application for breathing gymnastics.
This application consists of pre-defined breathing programs and levels, allowing users to select exercises based on their needs.
Figure \ref{fig:nafas-cycles} provides an overview of a program cycle, beginning with an inhale and progressing through multiple cycles, each with time durations specified in the program.

\textbf{Parameters.}
We use eight parameters for each program to control the effect on the respiratory system.
Program cycles are executed sequentially and repeated based on those parameters.
Each pre-defined program in \nafas includes unique ratios for the ``inhale,'' ``retain,'' ``exhale,'' and ``sustain.''
\nafas parameters are as follows:
\begin{enumerate}
\setcounter{enumi}{-1}
    \item \textbf{Preparation Time}: The time allowed before the breathing exercise begins, giving users a moment to get ready.
    \item \textbf{Inhale}: The time duration for drawing in air to oxygenate the body and cause lung expansion.
    \item \textbf{Retain}: The time duration for holding the breath after inhalation to enhance lung capacity.
    \item \textbf{Exhale}: The time duration for releasing breath, emptying the lung from the air.
    \item \textbf{Sustain}: The time duration for the breath-hold after exhalation.
    \item \textbf{Cycles}: Defines the total number of times the sequence of breathing steps is repeated.
    \item \textbf{Unit}: Determines the multiplication unit for time. For example, \verb|unit=3| would multiply a \verb|1:0:3:0| cycle by 3, yielding 3 seconds of inhale and 9 seconds of exhale.
    \item \textbf{Level}: Specifies the intensity or difficulty of the breathing exercise. 
\end{enumerate} 

\textbf{Programs.} Users can choose from 13 predefined programs such as ``Clear Mind,'' ``Relax,'' ``Anti-Stress,'' and ``Cigarette Replace.'' Each program is tailored to support specific states and goals, whether calming the mind, enhancing focus, or reducing stress.
You can see their description and parameters in Table \ref{tab:nafas_programs}.

The CLI design makes the tool easily accessible to developers and users comfortable working within the terminal. This approach enables users to perform breathing exercises during short breaks without disrupting workflow.

\section{Discussion}
Now, we will discuss some limitations to our work and future work that can enhance it.

% \subsection{Limitations}
While \nafas provides an integrated solution for performing breathing exercises within the command-line interface (CLI), it currently has some limitations that could impact its utility for a broader range of users.
Firstly, \nafas offers a fixed set of pre-configured programs across three levels without customization options. Although these pre-set programs cover common breathing techniques, they may not meet the specific preferences or needs of all users, particularly those who may have personalized breathing practices or advanced training. 

Furthermore, \nafas does not currently include any tracking or feedback mechanisms, so users cannot view their progress or evaluate improvements in their breathing routines over time.
For individuals who benefit from habit-tracking and progress visualization as a motivator, this absence may reduce long-term engagement with the application.

% \subsection{Future Directions}
Looking ahead, several enhancements are planned for future versions of \nafas to increase flexibility, personalization, and user engagement. One key improvement is the introduction of program customization, allowing users to create and modify their breathing routines based on personal preferences or specific goals. This feature will make \nafas accessible to a broader audience, including those with specialized health needs or breathing practices.

Additionally, \nafas can integrate with productivity tools like task managers or time-tracking software, prompting users to take breathing breaks based on their work intensity or schedule.
Another planned feature is adaptive difficulty, where \nafas adjust the intensity or duration of exercises based on user feedback or performance, making the tool more responsive to individual needs.
Finally, introducing tracking and analytics will allow users to monitor their progress.
These improvements will make \nafas a more versatile and user-friendly tool for developers and intensive computer users seeking accessible mindfulness practices.

\section{Conclusion}
This paper presents \nafas, a command-line application that provides breathing exercises tailored for computer users. Integrating these exercises directly into the terminal environment, \nafas offers a unique, non-intrusive solution to enhance focus and well-being. Future enhancements, including customizable programs and integration with productivity tools, will refine \nafas's role in supporting mental and physical health for intensive computer users.

% \newpage
% \onecolumn
\bibliography{main}
\bibliographystyle{acm}
% \twocolumn

% \section*{APPENDIX}

\end{document}